\documentclass[floatfix,aps,twocolumn,10pt,superscriptaddress,prapplied,amsmath,amssymb,showkeys]{revtex4-2}

\usepackage{graphicx}
\usepackage{dcolumn}
\usepackage{bm}
\usepackage{acronym}
\usepackage[]{subfig}
\usepackage[percent]{overpic}
\usepackage[hidelinks]{hyperref}
\usepackage{cleveref}
\usepackage{soul}
\usepackage[normalem]{ulem}
\usepackage{tikz}

\crefname{section}{Sec.}{Secs.}
\crefname{figure}{Fig.}{Figs.}
\crefname{equation}{equation}{equations}

\begin{document}

\title{Machine-learning based high-bandwidth magnetic sensing}

\author{Galya Haim}

\affiliation{Institute of Applied Physics, Hebrew University, Jerusalem 91904, Israel}
\affiliation{School of Physics, The University of Melbourne, Parkville, Victoria 3010, Australia}

\author{Stefano Martina}
\thanks{Corresponding author}
\affiliation{Dept. of Physics and Astronomy, University of Florence, via Sansone 1, I-50019 Sesto Fiorentino (FI), Italy}
\affiliation{European Laboratory for Non-Linear Spectroscopy (LENS), University of Florence, via N. Carrara 1, I-50019 Sesto Fiorentino (FI), Italy.}

\author{John Howell} 
\affiliation{Racah Institute of Physics, Hebrew University, Jerusalem 91904, Israel}

\author{Nir Bar-Gill} 
\affiliation{Institute of Applied Physics, Hebrew University, Jerusalem 91904, Israel}
\affiliation{Racah Institute of Physics, Hebrew University, Jerusalem 91904, Israel}

\author{Filippo Caruso}
\affiliation{Dept. of Physics and Astronomy, University of Florence, via Sansone 1, I-50019 Sesto Fiorentino (FI), Italy}
\affiliation{European Laboratory for Non-Linear Spectroscopy (LENS), University of Florence, via N. Carrara 1, I-50019 Sesto Fiorentino (FI), Italy.}
\affiliation{Istituto Nazionale di Ottica del Consiglio Nazionale delle Ricerche (CNR-INO), I-50019 Sesto Fiorentino (FI), Italy}

\collaboration{galya.haim@mail.huji.ac.il, stefano.martina@unifi.it, john.howell@mail.huji.ac.il, bargill@phys.huji.ac.il, filippo.caruso@unifi.it}

\begin{abstract}
Recent years have seen significant growth of quantum technologies, and specifically quantum sensing, both in terms of the capabilities of advanced platforms and their applications. One of the leading platforms in this context is nitrogen-vacancy (NV) color centers in diamond, providing versatile, high-sensitivity, and high-spatial-resolution magnetic sensing. Nevertheless, current schemes for spin resonance magnetic sensing (as applied by NV quantum sensing) suffer from tradeoffs associated with sensitivity, dynamic range, and bandwidth. Here we address this issue, and implement machine learning tools to enhance NV magnetic sensing in terms of the sensitivity/bandwidth tradeoff in large dynamic range scenarios. Our results indicate a potential reduction of required data points by at least a factor of 3, while maintaining the current error level. 
Our results promote quantum machine learning protocols for sensing applications towards more feasible and efficient quantum technologies.

\end{abstract} 
\maketitle

\acrodef{ml}[ML]{Machine Learning}
\acrodef{mlp}[MLP]{Multilayer Perceptron}
\acrodef{cnn}[CNN]{Convolutional Neural Network}
\acrodef{mse}[MSE]{Mean Squared Error}
\acrodef{mae}[MAE]{Mean Absolute Error}
\acrodef{mre}[MRE]{Mean Relative Error}
\acrodef{esr}[ESR]{Electron Spin Resonance}
\acrodef{odmr}[ODMR]{Optically Detected Magnetic Resonance}
\acrodef{nv}[NV]{Nitrogen Vacancy}
\acrodef{mw}[MW]{Microwave}
\acrodef{snr}[SNR]{Signal to Noise Ratio}
\acrodef{knn}[KNN]{K-Nearest Neighbor}
\acrodef{svm}[SVM]{Support Vector Machine}
\acrodef{cnn}[CNN]{Convolutional Neural Networks}
\acrodef{rnn}[RNN]{Recurrent Neural Networks}
\acrodef{llm}[LLM]{Large Language Models}
\acrodef{gan}[GAN]{Generative Adversarial Networks}
\acrodef{qml}[QML]{Quantum Machine Learning}
\acrodef{relu}[ReLU]{Rectified Linear Unit}

\section{Introduction}
Over the past decade, quantum technologies have emerged as an important platform relevant for a broad range of fields, such as quantum communications and quantum sensing. These advances have been driven by the development of experimental realizations exhibiting needed and useful properties.

In the context of quantum sensing, one of the leading systems is based on \ac{nv} color centers in diamond~\cite{PhysRevB.85.205203}, which provide a versatile platform for diverse quantum sensing, notably magnetic sensing~\cite{Grinolds2013-jr}.
\acp{nv} have found important applications in magnetic sensing, covering paleomagnetometry~\cite{19197}, biosensing~\cite{26099}
nuclear magnetic resonance~\cite{doi:10.1126/science.1231540}
and more. 

\begin{figure*}[tbh]
    \centering
    \subfloat[\label{fig:scheme-setup}]{%
        \centering
        \includegraphics[width=0.25\linewidth]{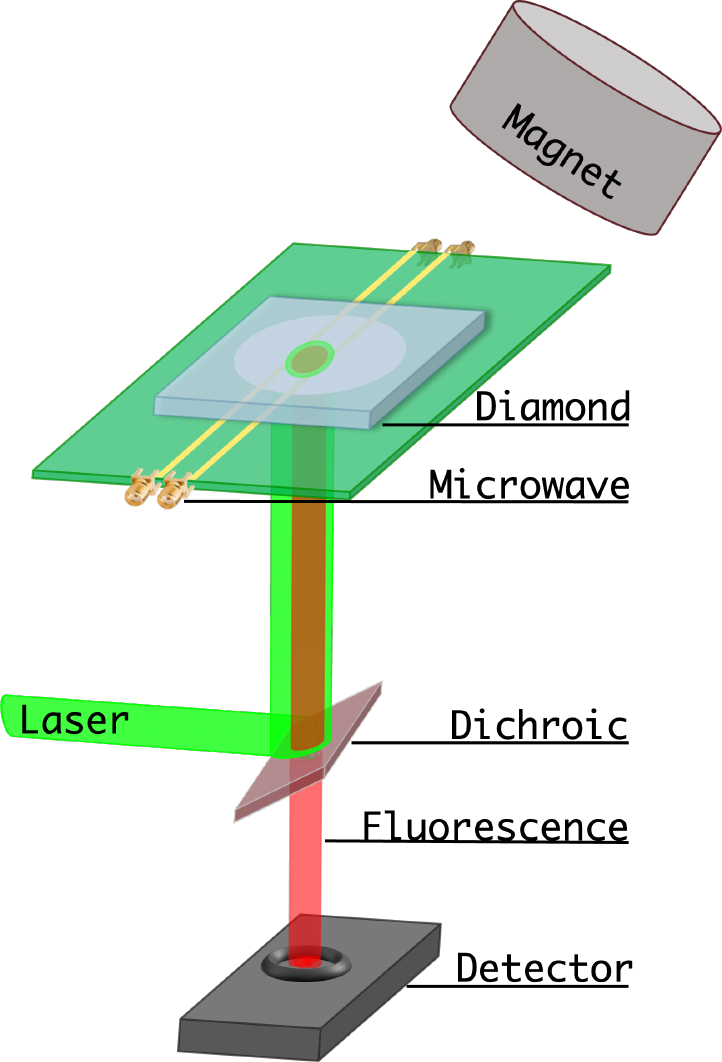}%
    }\hspace{2cm}%
    {
        \captionsetup[subfigure]{labelformat=empty}
        \subfloat[][]{\label{fig:scheme-nv}}
    }%
    \subfloat[\label{fig:scheme-level}]{%
        \centering
        \begin{overpic}[width=0.35\linewidth]{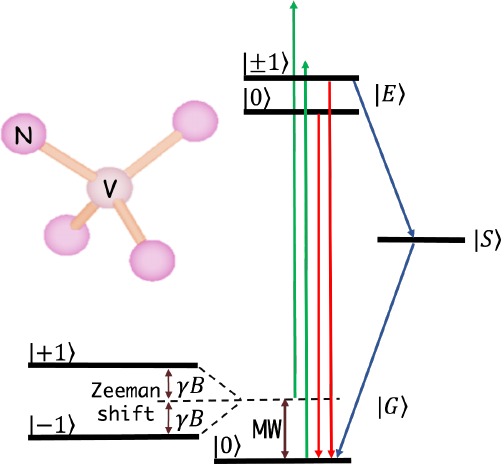}%
            \put (16,32) {(b)}%
        \end{overpic}%
    }
    
    \subfloat[\label{fig:scheme-esr}]{%
        \centering
        \includegraphics[width=0.35\linewidth]{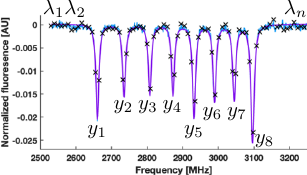}%
    }\hspace{1cm}
    \subfloat[\label{fig:scheme-net}]{%
        \centering
        \includegraphics[width=0.35\linewidth]{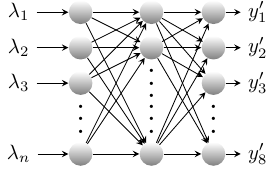}%
    }
    \caption{\protect\subref{fig:scheme-setup} Experimental setup schematic; MW pulses are used to identify the spin resonances of the \ac{nv} center within the diamond sample. The resonance frequencies are related to the external magnetic field orientation and intensity through the Zeeman shift. Green excitation laser illuminates the diamond and fluorescence from the \ac{nv} centers is collected and read by a detector. \protect\subref{fig:scheme-nv} The \ac{nv} can appear in 4 possible orientations within the diamond lattice. \protect\subref{fig:scheme-level} Schematic of the \ac{nv}'s energy levels. Under an external magnetic field, the degeneracy between the $m_s = \pm 1$ energy levels of the ground state is lifted. \ac{mw} pulses drive the transitions between $m_s =0$ and $m_s = \pm 1$. The ground state is excited with a green laser (532 nm) and can decay via two pathways: one is non-radiative and non-spin-conserving through the singlet state, while the other is spin-conserving and emits red fluorescence in the range of 650-800 nm. The $m_s = 0$ state decays mostly radiatively, while $m_s = \pm 1$ includes a significant non-radiative decay through the singlet state into the ground spin state $m_s = 0$. \protect\subref{fig:scheme-esr} \acs{esr} spectrum, with measured data in blue and the corresponding fit in purple (to a set of 8 Lorentzian functions). In black, a subset of the data, i.e $\lambda_i$. The resonance frequencies are marked with $y_i$. \protect\subref{fig:scheme-net} Schematic of the neural network, the inputs $\lambda_i$ are the measured data points and the outputs are the predictions $y'_i$ of the 8 resonance 
    frequencies $y_i$.}
    \label{fig:scheme}
\end{figure*}

Quantum sensing with \acp{nv} is realized through the illustrated spin resonance measurement techniques. Although this approach achieves quantitative vectorial information with high sensitivity and spatial resolution, it suffers from a trade-off between sensitivity and bandwidth, specifically in the high-dynamic range regime. In fact, working with small fields (small-dynamic range) enables an optimal sensing strategy, which relies on precise measurements at a predetermined high-sensitivity point (the point of maximal signal gradient). However, this is not possible in the regime of large-dynamic range signals.
As shown in \cref{fig:scheme}, the standard approach to quantitatively measure vectorial magnetic fields using \acp{nv} is essentially a common spin resonance measurement termed \ac{esr}, usually performed through optical readout for \acp{nv}, referred to as \ac{odmr}, using an experimental setup schematically depicted in \cref{fig:scheme-setup} \cite{Maze2008-lt}. 

The orientation of NV defects within the crystal lattice (\cref{fig:scheme-nv}) and its energy level structure (\cref{fig:scheme-level}), enable the optical detection of NV spin resonances, realized through the combined application of green-light excitation and microwave (MW) irradiation, while detecting the NV spin-dependent red fluorescence.
As depicted in \cref{fig:scheme-esr}, a full \ac{esr} spectrum of an \ac{nv} ensemble consists of 8 resonances, representing 2 resonant frequencies for each of the 4 crystallographic orientations the \ac{nv} can take in a single crystal diamond sample (\cref{fig:scheme-nv}). 
Basically, this measurement amounts to identifying resonance positions in the frequency space.
Once obtained, a full vectorial magnetic field can be extracted. 
Further details regarding the \ac{nv} system, the experimental setup (\cref{fig:scheme-setup,fig:scheme-level}) and the \ac{odmr} measurement scheme can be found in \cref{sec:methods}.

Identifying the resonance positions requires scanning the signal over the frequency space, often referred to as raster scanning. This scan must cover the entire relevant frequency range, determined by the desired dynamic range. At higher fields, the Zeeman splitting increases, resulting in larger separations between resonances. This enables a wider dynamic range, allowing the system to measure both small and large magnetic fields without overlapping resonances. The scan resolution defined by the number of data points within the frequency range, depends on the resonance widths and \ac{snr}. 
There is a minimal scan resolution required to successfully retrieve  the resonance frequencies, and in general the measurement error depends on these various parameters (linewidth, \ac{snr}, scan resolution) as detailed in \cref{sec:methods}.

The sensitivity for an \ac{odmr} measurement is defined to be $\eta = (\delta B*\sqrt{T})$ ; or equivalently in terms of frequency,  $\eta = (\delta \nu *\sqrt{T})/\gamma$, where $\delta B$ is the uncertainty in magnetic field $\delta \nu$ is the error in frequency, T is the measurement time which corresponds to the number of data points, averaging and \ac{snr}, and $\gamma$ is the gyromagnetic ratio of the \ac{nv}~\cite{Pham2013-da}. 
A shorter measurement time, while maintaining the same SNR, improves sensitivity and enables the measurement of time-varying signals.
Reducing measurement time could be achieved by improving the experimental setup, i.e., having higher \ac{snr} or better \ac{nv} properties. However, it is not less important to consider the optimal number of data points in a measurement. 

In this paper, we propose an \ac{ml} algorithm as a promising approach to address the current trade-off. We demonstrate that training an appropriate neural network using a combination of real and simulated data enables an improvement in measurement bandwidth for a given sensitivity goal, in the large dynamic range scenario.

\ac{ml} is a research field that deals with the development of artificial intelligence methods that learn from data. Several approaches were developed to solve the two main supervised \ac{ml} tasks of classification, i.e.\ to predict a categorical value associated to some predictors (called features in the \ac{ml} jargon), and regression where the predicted value is continuous~\cite{hastie2009elements}.
Our work lies within the emerging field of \ac{qml}, positioned at the intersection of \ac{ml} quantum physics and quantum computing~\cite{biamonte2017}.
In general, \ac{qml} encompasses various approaches, including the application of classical \ac{ml} models to quantum systems, for example, in noise analysis of quantum devices~\cite{martina2022learning}, or
the implementation of \ac{ml} models on quantum devices,
for example, quantum generative models~\cite{parigi2024diffusion}.

Neural networks are powerful \ac{ml} models constituted by interconnected layers of artificial neurons that are trained to minimize a loss function~\cite{Goodfellow2016}.
Recently, thanks to the advances in computational power and the growing availability of large datasets, neural networks have outperformed other \ac{ml} algorithms, becoming a general-purpose tool for either clustering, classification, regression~\cite{bishop2006pattern} and generative~\cite{paiano2024transfer} tasks. In fact they are capable of universal function approximation~\cite{hanin19} and easily adapt to different scenarios. Their flexibility allows them to serve as the underlying structure in various architectures and tasks.
In our work we adopt a classical feed-forward architecture, commonly known as \ac{mlp}~\cite{lecun2015deep},
as illustrated in \cref{fig:scheme-net}. The network is trained using gradient descent to minimize a loss function between the prediction and the desired outputs. More details can be found in \cref{sec:machineLearning}.
Such models have already been used with \ac{nv} centers for sensing~\cite{tsukamoto2022accurate},
and for noise spectroscopy~\cite{Martina_2023}. These approaches hold promise for various practical applications, such as in the study of integrated circuits~\cite{ashok2022hardware}
and medical diagnostics~\cite{tehlan2024magnetization}.  
In our work, we apply neural networks to the full reconstruction of an external magnetic field using an \ac{nv} ensemble.
We analyze the network's performance for a decreasing number of measurement points, while exploring different training dataset sizes, and altering noise and lineshape conditions.
We compare these results to similar analysis done for raster scans and find that the network has advantages over raster scanning. With the right training, the network can be insensitive to variations in noise and lineshape. Moreover, the network's measurement error is more robust and scales better for fewer data points.

\section{Methods}\label{sec:methods}
\subsection{Experimental setup and data acquisition}
The ground state of \ac{nv} centers (\cref{fig:scheme-level}) is an effective two-level quantum system.
Under green laser excitation, it is possible to initialize the spin to the $m_s=0$ ground state. In detail, the population occupying $m_s = 0$ would reach the excited state manifold and decay back to the ground state $m_s = 0$, emitting red photons. The population occupying $m_s = \pm1$ is more likely to decay through the singlet state, to $m_s = 0$ in a non-radiative way. 
Within the ground state, spin manipulation is possible with resonant \ac{mw} pulses, population transfer to $m_s =\pm 1 $ would lead to a drop in measured fluorescence. 

In the presence of an external magnetic field, degeneracy is lifted off the $m_s =\pm 1 $ due to Zeeman shift, the shift is given by $\gamma B_\parallel$, where $\gamma $ is the gyromagnetic ratio of the \ac{nv} and $B_\parallel$ is the external magnetic field component parallel to the \ac{nv} axis. In the diamond lattice, there are 4 possible crystallographic orientations the \ac{nv} can take (\cref{fig:scheme-nv}), and so, in the presence of a magnetic field that is not aligned with any of the orientations, there will be 8 resonance frequencies as shown in \cref{fig:scheme-esr}, 2 for each orientation.
Once these frequencies are known, the vectorial magnetic field can be calculated.

Ideally, an \ac{esr} measurement would have the smallest number of points for which all information about the magnetic field can be obtained from the data, without increasing measurement error, which means lower sensitivity.
As can be expected, the error increases as the number of data points decreases. 
Below a certain number of data points, raster scanning is simply not possible as not all 8 Lorentzians show in a data sample, this number varies and depends on the measurement bandwidth as well as on \ac{snr}. 
As mentioned in the introduction, the sensitivity is defined to be $\eta = (\delta \nu *\sqrt{T})/\gamma$. 
To account for the statistical success of the raster in obtaining 8 Lorentzians, we normalize the measurement time, T, by P - the success probability, and so the normalized sensitivity is now $\eta = (\delta \nu *\sqrt{T/P})/\gamma$.
Based on that, we define the normalized error to be $\delta \nu /\sqrt{P}$.

\begin{figure}[htb]
    \centering
    \subfloat[\label{fig:rasterwidth}]{%
        \centering
        \includegraphics{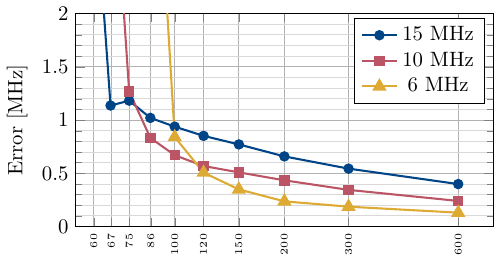}
    }

    \subfloat[\label{fig:rastersnr}]{%
        \centering
        \includegraphics{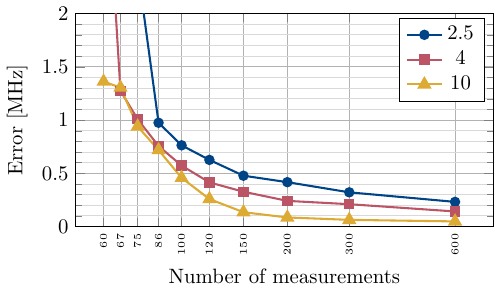}
    }
    \caption{The raster subsampling error (not normalized) is shown as a function of the number of data points, averaged over 100 data samples. 
    In \protect\subref{fig:rasterwidth}, simulated data with a constant \ac{snr} of 4 and changing Lorentzian widths: 6 (yellow triangles), 10 (red squares) and 15 (blue circles) MHz.
    In \protect\subref{fig:rastersnr}, simulated data with a constant width of 10 MHz, and different \acp{snr}: 10 (yellow triangles), 4 (red squares) and 2.5 (blue circles).}
    \label{fig:Rasternoisewidth}
\end{figure}

\Cref{fig:Rasternoisewidth} depicts the average error of 100 subsampled simulated raster scans for different \acp{snr} and Lorentzian widths.
For a constant \ac{snr} of 4 (\cref{fig:rasterwidth}), error is lower for narrow Lorentzians. However, the wider Lorentzians are, they can be identified with a smaller number of data points. 
\cref{fig:rastersnr} depicts the averaged error of 100 subsampled simulated raster scans, all of a constant Lorentzian width of 10 MHz, but varying \ac{snr}. As expected, the error increases as \ac{snr} decreases. 

\ac{esr} raster scans were done in an epi-illumination wide field home built microscope.
Green laser (lighthouse, Sprout), illuminates the diamond sample through an objective, and fluorescence is collected by a camera (Andor, Neo). Microwaves (generated by Windfreak, SynthHD) are delivered to the sample with a custom made waveguide (\cref{fig:scheme-setup}).

\subsection{Machine learning model}\label{sec:machineLearning}
The \ac{ml} model depicted in~\cref{fig:scheme-net} is employed to predict the resonance positions $y_i$ in the \ac{esr} spectrum. 
The dimensions of the input and output layers are determined by the task, where the input corresponds to the \ac{esr} data points $\lambda_1,\dots,\lambda_n$, and the output corresponds to the 8 resonance positions. The number of layers and their size are treated as hyperparameters to be optimized, and to keep the process simple, the number of neurons is the same across the hidden layers. The task is framed as a regression problem with all the inputs and outputs normalized to fall between 0 and 1. The activation functions for all layers are \ac{relu}, defined by $\max(0,x)$, except for the sigmoid in the output layer.

In detail, we employ an \ac{mlp} model composed of fully connected layers, each containing a certain number of artificial neurons. A single artificial neuron returns the scalar
\begin{equation*}
\hat{y} \equiv f(\bm{w}^T\cdot\bm{x}+b)
\end{equation*}
where $f:\mathbb{R}\rightarrow\mathbb{R}$ is a nonlinear function applied to the weighted sum of the inputs $\bm{x}\in\mathbb{R}$ with the weights $\bm{w}$ plus the bias term $b$. In general, an \ac{mlp} with $L$ layers is defined by the relation
\begin{equation}\label{eq:mlp}
    \bm{h}_\ell = f_\ell\left(W_\ell^{T}\bm{h}_{\ell-1}+\bm{b}_\ell\right),
\end{equation}
for $\ell=1,\dots,L$, where $\bm{h}_\ell\in\mathbb{R}^{k_\ell}$ is the vector of $k_\ell$ outputs of the $\ell$-th layer, and $\bm{h}_{\ell-1}\in\mathbb{R}^{k_{\ell-1}}$ is the vector of the $k_{\ell -1}$ outputs from the previous layer. The matrices $W_\ell\in\mathbb{R}^{k_{\ell-1}\times k_\ell}$ and $\bm{b}\in\mathbb{R}^{k_\ell}$ represent the weights and biases of all neurons in layer $\ell$. 
The values of $L$ and $k$ are optimized with a dedicated framework~\cite{liaw2018tune} along with other training hyperparameters. Specifically, the \emph{learning rate}, the \emph{batch size} and the amount of regularization with \emph{dropout}~\cite{Baldi2013dropout} and \emph{weight decay}. 
The neural networks are trained with \emph{gradient descent} to minimize the error of the predictions $\bm{y}'$ respect to the ground truth $\bm{y}$ in the form of the \ac{mse} loss:
\begin{equation*}
    \mathcal{L}(\bm{y},\hat{\bm{y}})=\frac{1}{8n}\sum_{i=1}^n\sum_{j=1}^8\left(y_{i,j}- \hat{y}_{i,j}\right)^2.
\end{equation*}
A generic gradient descent method aims to find the parameters $\bm{\theta}$ of the neural network ($W_\ell^T$ and $\bm{b}_\ell$ in \cref{eq:mlp} for all $\ell$). The gradient descent method is implemented iteratively by calculating 
\begin{equation}\label{eq:sdg}
    \bm{\theta}_{t+1}=\bm{\theta}_t-\eta\nabla_{\bm{\theta}}\mathcal{L}(\bm{y}|_{B_t},\hat{\bm{y}}|_{B_t}),
\end{equation}
where $\eta$ is the learning rate that determines the step length, $\nabla_{\bm{\theta}}$ is the gradient of the loss surface in $\bm{\theta}$ and the notation $\cdot|_{B_t}$ indicates that we use a subset $B_t$ of the training dataset called \emph{minibatch}, whose dimension is determined by the batch size. The entire training dataset is used iteratively, batch by batch, to perform the training steps. A full pass through all the data is called an \emph{epoch}.
Specifically, we train the models using Adam~\cite{kingma2014adam}, an optimizer that implements gradient descent with \emph{adaptive momentum}. Momentum is a technique that accelerates gradient descent by incorporating the weighted averages of gradients from the previous steps during the process. In Adam the weights of the averages are dynamically adjusted during the training.

The prediction error for the neural networks is quantified by the \ac{mae}:
\begin{equation*}
    MAE(\bm{y}, \hat{\bm{y}}) = \frac{1}{8n} \sum_{i=1}^n \sum_{j=1}^8 \left|y_{i,j} - \hat{y}_{i,j}\right|,
\end{equation*}
where $y_{i,j}$ is the ground truth value and $\hat{y}_{i,j}$ is the predicted value for the $j$-th resonance of the $i$-th sample. Regarding the Lorentzian fitting procedure, we integrate the prediction confidence into the error calculation, thus we define the \ac{mae} for that case as:
\begin{equation*}
    MAE(\bm{y}, \hat{\bm{y}},\bm{c}) = \frac{1}{8n} \sum_{i=1}^n \sum_{j=1}^8 \sqrt{\left(y_{i,j} - \hat{y}_{i,j}\right)^2+c_{i,j}^2},
\end{equation*}
where $y_{i,j}$ and $\hat{y}_{i,j}$ are the same as before and $c_{i,j}$ is the confidence for the prediction of the $j$-th resonance of the $i$-th data sample.

All models are trained for a maximum of $100$ epochs. In the end we keep the parameters for the model that have the minimum \ac{mae} error on the validation set and we use it to report the results.

Hyperparameter optimization is a well-known challenge in \ac{ml}. It is often resource-intensive, as the choice of hyperparameters is typically task-specific, with no universal method for determining their values. In our work, this burdensome optimization is alleviated through the use of the \emph{Ray Tune} platform~\cite{liaw2018tune} that parallelizes the runs across different hyperparameter combinations and employs a Bayesian optimization technique called \emph{Hyperopt}~\cite{hyperopt}. Hyperopt uses Tree-structured \emph{Parzen} estimators~\cite{bergstra2011algorithms} to suggest the most probable best combination for the hyperparameters.
Moreover, to accelerate the process, the \emph{ASHA} scheduler~\cite{asha} was adopted to terminate the least promising trials before completing their full training procedure. 

The ranges that were used for hyperparameters optimization are as follows: 
$\eta\in\{10^{-2},10^{-3},10^{-4}\}$ for the learning rate; $|B|\in\{2,4,8,16,32\}$ for the batch size; the dropout is within $\{0,0.2,0.5\}$, and the weight decay within $\{0, 10^{-6}, 10^{-5}, 10^{-4}, 10^{-3}\}$.
$\mathcal{LU}(a,b)$ is a uniform distribution of integers within the range $(a,b)$ in the $\log_{10}$ domain (called \emph{lograndint} within \emph{Ray Tune}), used to define the range for the number of layers $L\sim \mathcal{LU}(1,32)$ and their size $k\sim \mathcal{LU}(1,1024)$.

The codes used for the implementation and training of the neural networks and for the generation of simulated data are available on the GitHub repository~\footnote{\url{https://github.com/trianam/machineLearningMagneticSensing}}.

\section{Results}\label{sec:NNtrainingAndEvaluation}
96 full raster scans were measured in an epi-illumination wide field setup (see \cref{fig:scheme-setup} and \cref{sec:methods} for details), each one under a different applied magnetic field, corresponding to a change in the resonances locations within the frequency window. These were measured through standard ODMR.

\begin{figure*}[htb]
    \centering
    \subfloat[\label{fig:nnSynth_2}]{%
        \centering
        \includegraphics{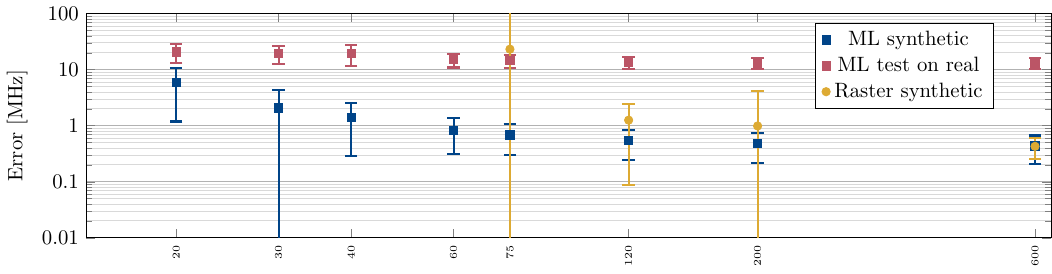}
    }
    
    \subfloat[\label{fig:nnlfIntSynth10k}]{%
        \centering
        \includegraphics{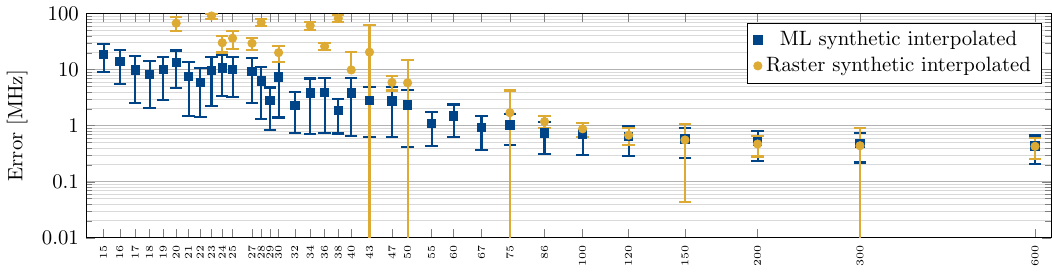}
    }
    
    \subfloat[\label{fig:realdata}]{%
        \centering
        \includegraphics{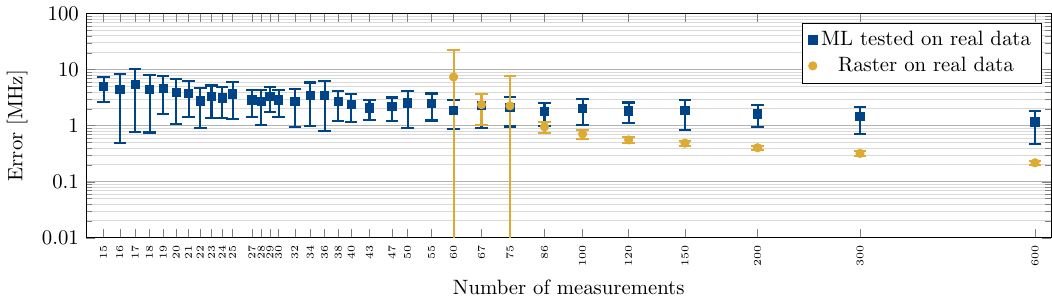}
    }
    
    \caption{Comparison between the error of neural networks and the normalized error of raster scans for varying numbers of data points. The plot, shown on a logarithmic scale, reports the average error values and their standard deviations across different number of measurements. Each value for the \ac{ml} models has a corresponding value from raster scanning. For raster scans, data points are omitted when the 8 peaks cannot be identified or when the resulting error exceeds a maximum value of $10^6$ MHz. In \protect\subref{fig:nnSynth_2} and \protect\subref{fig:nnlfIntSynth10k}, \ac{ml} networks (blue squares) were trained with $10\,000$ synthetic samples, validated on $2\,000$ samples and tested on an additional $2\,000$ samples to report the results. The same 
2,000 test samples were used to calculate the normalized error of raster scanning (yellow circles). In red squares the same networks were tested  on 46 real data samples. In \protect\subref{fig:nnSynth_2} the dataset was subsampled to the specified number of data points, and a different networks were trained and tested for each subsampling ratio. Similarly, the raster scan was calculated directly on the subsampled data. In \protect\subref{fig:nnlfIntSynth10k} only one network was trained with the $600$ data points of the full dataset. For testing, subsampled data was linearly interpolated to upsample it to the full size. The same interpolation procedure was applied to the raster scan. 
   \protect\subref{fig:realdata} The networks (blue squares) were trained with subsampled datasets, similar to those in \protect\subref{fig:nnSynth_2}. However, the training dataset contained $1\,000$ synthetic samples and 50 real data samples, while validation was performed using the same 46 real data samples as in \protect\subref{fig:nnSynth_2}. The same 46 samples were used for the raster scan, and their average normalized error is shown in yellow circles. In all 3 cases the simulated data had random SNR in the range [2.5, 10], and a Lorentzian width of 10 MHz. The real data had SNR within the same range, and an average Lorentzian width of 10 MHz with small variations between measurements and\ or the 8 resonances.}
    \label{fig:10k}
\end{figure*}

Synthetic data was generated based on a simplified \ac{nv} Hamiltonian~\cite{PhysRevB.85.205203} considering the Zeeman shift:
magnetic fields at various angles were projected on the 4 \ac{nv} orientations, determining the resonance frequencies. Resonance widths, contrast, and Gaussian noise were chosen to mimic the lineshape of the measured data, producing full \ac{esr} spectra with 6 or 8 Lorentzians, similar to the one depicted in~\cref{fig:scheme-esr}.

The neural network's input layer dimension ($\lambda_n$ in ~\cref{fig:scheme-net}) is determined by the number of data points in an \ac{esr} spectrum and therefore was changed as networks were trained for different subsamplings: starting from the full 600-point spectrum, we subsampled by taking every other point, every third point and so on.
The output layer always consists of 8 values that predict the central frequencies of the 8 Lorentzians ({$y\;'_i$}). 
These predictions were then compared to the values extracted from the full \ac{esr} spectrum, which was defined to be the ground truth. The error was defined to be the averaged absolute value of the differences.

To accomplish a fair comparison, a similar process was performed for raster scans: the resonances extracted from subsampled data, through standard Lorentzian curve fitting, were compared to the ones extracted from the full spectrum. The error was defined in the same way. 
It is important to note that the subsampled datasets were compared to the full data from which they were derived, and the error in the full scan is the fit error.

Some of the simulated data samples had overlapping or partly overlapping Lorentzians, for which the centers were sometimes too close to be distinguished. Note that for these data samples the neural network still outputs 8 values. However, trying to identify 8 resonances using a raster scan, especially with lower \ac{snr}, would often fail. Raster scans also fail when heavily subsampled, as one or more of the Lorentzians are no longer recognizable. To account for this statistical trait of the raster scans, we normalize the error by the square root of the success probability, essentially following the definition of the sensitivity and considering the success probability in terms of time (explained in detail in \cref{sec:methods}).

\cref{fig:nnSynth_2} depicts the \ac{ml} error (blue squares) for networks that were trained with $10\,000$ samples and validated with $2\,000$ (more details on the model are available in \cref{sec:machineLearning}, and further details on the size of the data set are available below). As previously described, each network was trained on the relevant subsampled data. Yellow circles depict the normalized error of subsampled raster scans, averaged over the same $2\,000$ samples that were used for the network's testing. 
The raster scan error is distinctly high, due to the above-mentioned normalization of the error with the success probability, since in some of the data samples it was not possible to identify 8 Lorentzians. 
The \ac{ml} model exhibits better results compared to raster scanning with an error reduced by more than 400 kHz, except for the case of full length data. Remarkably, with only $10\%$ of the data points the \ac{ml} error is still below $1$ MHz. In addition, the \ac{ml} error has better scaling as a function of the number of data points. The model maintains similar normalized error for 200 data points as it does for 600 data points, which gives a factor 3 on measurement time preserving the measurement error. Additionally, for 120 data points, the mean \ac{ml} error falls within $0.1$ MHz of the raster scan error obtained using 600 data points. The same networks, trained on simulated data, are evaluated on 46 real data samples (red squares). The results consistently show higher errors compared to testing on simulated data, indicating overfitting during training, as illustrated in \cref{fig:realTest} of the supplementary material.

In \cref{fig:nnlfIntSynth10k} we present an alternative training scheme: in this case, just one network was trained and only on the full length data with $10\,000$ samples and validated with $2\,000$ samples. 
To comply with the structure of the \ac{mlp}, the network was tested on $2\,000$ subsampled dataset, which was first linearly interpolated back to its full length before being introduced to the network. The averaged error for these tests is presented in the plot (blue squares). 
The same $2\,000$ samples were used to test the raster scan (yellow circles), these were also sub-sampled, interpolated and then assessed.
The interpolation surprisingly reduces the error for raster scan. Even though it introduces additional noise to the data, the fitting works significantly better. For a higher number of points, the raster's error and standard deviation are very similar to the network. However, below 120 data points the \ac{ml} model shows better results compared to the raster scan.
Interpolating data saves on training time, since one network fits all datasets, but at the cost of a less favorable performance. When a model is trained on the full 600-points data, it learns patterns that can be distorted with the subsampling and the subsequent interpolation.
For high numbers of points, the two methods are almost equivalent. Although, for lower numbers of points, starting at 120, a network trained for the specific number of data points gives a lower averaged error and standard deviation, as can be seen from the comparison of the two graphs in \cref{fig:nnSynth_2} and \cref{fig:nnlfIntSynth10k} (Also depicted in Supplementary \cref{fig:nnSynth}).

We note that, while not realized here, a hybrid approach might be considered and could be beneficial, wherein several networks could be trained on different subsampling sizes but not on all of them. In this case, the data to be analyzed could be interpolated to the closest available dimension and the corresponding trained network can be employed for the prediction. This can save training time while keeping the predictions less affected by the interpolation of the data.

\cref{fig:realdata} depicts the results of networks (blue squares) tested on real data, and the subsampling of real raster scans.
The networks were trained (as in \cref{fig:nnSynth_2})
for specific subsampling ratios, using a combination of $1\,000$ synthetic data samples along with 50 real data samples. They were tested on additional 46 real data samples, which were also used to calculate the raster's normalized error. 
The raster's results in \cref{fig:realdata}, are better than in \cref{fig:nnSynth_2} and \cref{fig:nnlfIntSynth10k}. This is mostly due to the fact that in all 46 data samples, the 8 Lorentzians were well separated. 
The network's error here still has a better scaling than the raster's error, however, for a high number of points the network error is about $1$ MHz worse than the raster, which is in agreement with training predictions that are further explained in the following. Moreover, the error observed in the test on real samples in \cref{fig:realdata} is one order of magnitude lower than that shown in \cref{fig:nnSynth_2}.

We now turn to a more detailed analysis, further examining the behaviour of the network with regards to other parameters: the size of the training dataset, \ac{snr} and Lorentzian widths.

\begin{figure}[htb]
    \centering
    \includegraphics{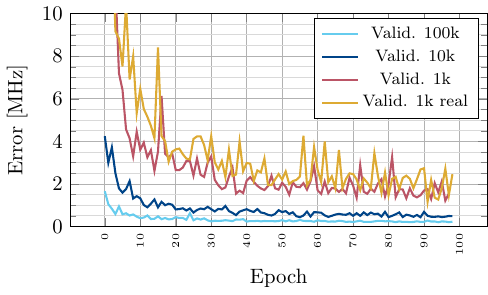}
    \caption{Evolution of error during the training of \ac{ml} models, calculated on the validation sets of different datasets. 
    In light blue (Valid. 100k): training with $100\,000$ synthetic samples and validated on $20\,000$ synthetic samples.
    In blue (Valid. 10k): training with $10\,000$ synthetic samples and validated on $2\,000$ synthetic samples. 
    In red (Valid. 1k): training with $1\,000$ synthetic samples and validated on 200 synthetic samples. 
    In yellow (Valid. 1k real): training with $1\,000$ synthetic samples plus 50 real samples, validated on 46 different real samples. }
    \label{fig:trainSynth}
\end{figure}

\Cref{fig:trainSynth} depicts the evolution of the validation error during the training of the neural network for 3 dataset sizes: in  light blue, the network was trained with $100\,000$ simulated samples and validated with $20\,000$ samples. in blue, the network was trained with $10\,000$ simulated samples and validated with $2\,000$ samples. In red, the network was trained with $1\,000$ simulated samples and validated with 200 simulated samples. In yellow, 50 real data samples were added to the training set, and the validation was performed on 46 real data samples. 
The plot shows that training with $1\,000$ data samples is insufficient, as the network trained with $10\,000$ data samples presents a significantly lower error. Furthermore, training with $100\,000$ data samples yields an additional improvement; however, it comes at the cost of significantly increased computational demands, and can be optimized.
This plot explains the difference between results depicted in \cref{fig:realdata} and \cref{fig:nnSynth_2} - respectively represented here by the yellow and blue curves, with a difference of $0.5-2$ MHz between them. 

\begin{figure*}[htb]
    \centering
    \subfloat[\label{fig:10mhz}]{%
        \centering
        \includegraphics{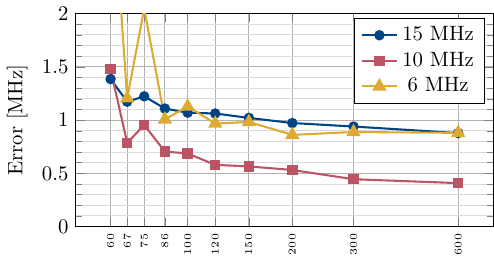}
    
    }\hfill
    \subfloat[\label{fig:widthrange}]{%
        \centering
        \includegraphics{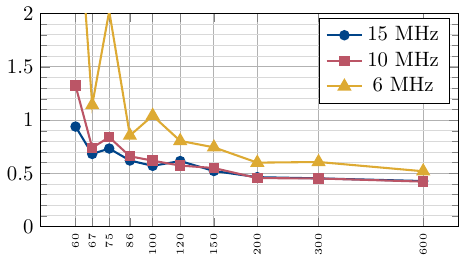}
    }

    \subfloat[\label{fig:noiseVarFix}]{%
        \centering
        \includegraphics{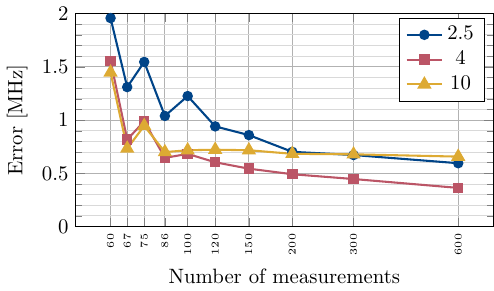}
    }\hfill%
    \subfloat[\label{fig:noiseVar}]{%
        \centering
        \includegraphics{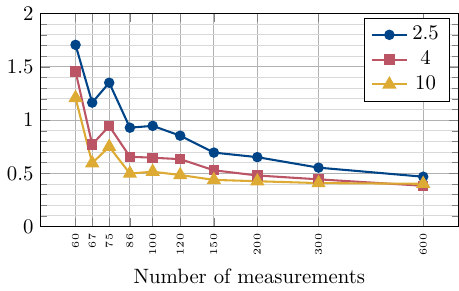}
    }
    \caption{\ac{ml} averaged error on simulated data for different number of data points,  illustrating the effect of training strategies to enhance the model's robustness to variations in the data. All models are trained on $10\,000$ samples with $600$ data points and tested on 3 different datasets of $100$ samples subsampled and successively interpolated to the full dimension. 
   In particular, in \protect\subref{fig:10mhz} the network is trained on samples having 10 MHz Lorentzian width and tested on 3 different datasets, each with a different width: 6 MHz (yellow triangles), 10 MHz (red squares), and 15 MHz (blue circles). In \protect\subref{fig:widthrange} the network is trained on samples with random width within the range $[5,16]$ MHz and tested on the same datasets as in \protect\subref{fig:10mhz}.
   \protect\subref{fig:noiseVarFix} depicts the averaged error for a model trained on samples with \ac{snr} value of 4 and tested on different samples with \ac{snr} values of 2.5 (blue circles), 4 (red squares), and 10 (yellow triangles). 
   In \protect\subref{fig:noiseVar}, the network was trained on samples with random \ac{snr} within a range of $[2.5, 10]$ and tested on the same datasets as in \protect\subref{fig:noiseVarFix}.}
    \label{fig:MLwidthtraining}
\end{figure*}

Two important parameters which characterize the resonances measured in ESR experiments are the \ac{snr} and Lorentzian width. These vary between experimental setups, diamond samples and even individual measurements due to different noise sources~\cite{Taylor2008-qs}. \cref{fig:MLwidthtraining} presents the average error of two networks: in \cref{fig:10mhz} the network was trained on a simulated dataset in which all $10\,000$ samples had a width of 10 MHz and subsequently tested on 3 datasets with $100$ data samples, each with different resonance widths: 6 (yellow), 10 (red), and 15 MHz (blue).  
When tested on $10$ MHz, the network's averaged error is 0.5 MHz lower than it is for 6 and 15 MHz.
In \cref{fig:widthrange} the network was trained on a different dataset of $10\,000$ samples with widths within a range from 5 to 16 MHz. It was then tested on the same 3 datasets as in \cref{fig:10mhz}. 
Training on a range of widths proved to be a more robust method. The errors for all the tests are comparable and the network is no longer sensitive to changes in width.
Similar behavior was observed for \ac{snr}. \cref{fig:noiseVarFix} depicts a network that was trained on data samples with \ac{snr} of 4, and tested on different samples with \ac{snr} 2.5, 4 and 10. \cref{fig:noiseVar} depicts the test results of a network that was trained on samples with \ac{snr} randomly sampled within the range $[2.5, 10]$. Again, the network that was trained on a range is more robust, and when training specifically, the variations show a higher error. The results in \cref{fig:MLwidthtraining} report the averages on $100$ test samples, for different number of measurements, using the interpolation training scheme.  

As described in detail in \cref{sec:methods}, Changes in noise and width also have an effect on subsampling raster scans. In particular, the results of the trained \ac{ml} models in \cref{fig:MLwidthtraining} can be compared with the raster scans in \cref{fig:Rasternoisewidth}. 
It is important to note that only the raster scan values that successfully identified all 8 resonances were included in \cref{fig:Rasternoisewidth}, excluding the normalized error. A comparison including the normalization is available in \cref{fig:LFwidthtraining} of the supplementary.

\section{Conclusion}
In this paper we employ \ac{ml} models to achieve high bandwidth measurements without compromising sensitivity in \ac{esr} measurements with \ac{nv} centers.
The neural networks exhibit an advantage over raster scans, they maintain the same error for down to fifth of the data points, while for the same sub sampling rate, the raster's error increases by about 800 KHz.
The \ac{ml} models still perform well with an error of less than $1$ MHz using only $10\%$ of data points. Moreover, they show an impressive ability to predict resonance locations in the presence of overlapping Lorentzians (degeneracy). To clarify, the training included data with 6 resonances instead of 8 and the network learned to correctly identify all 8 resonances for such data samples. We believe this point warrants further investigation to understand the limits of this feature.
We show the gain and flexibility of training networks in different ways: interpolating can save on training time while training multiple networks yields a lower error and standard deviation. Furthermore, we show that for a network to be robust, it needs to be trained on a sufficiently large dataset that includes data spanning a range of noise levels and linewidths.

We find that the network's performance will benefit from including in the training a range of snr, lineshape, and increasing the dataset size. In addition, we find it essential to have a simulation that accurately captures the physics of the \ac{nv} system, ensuring that the resulting synthetic data closely match the characteristics of the measured data.

The machine learning techniques applied to this quantum setting prove to be efficient, achieving an improved trade-off between high sensitivity and high dynamic range. 
They can be adaptively applied to measurements to achieve the desired result. E.g., reducing the measurement time by a at least factor 3, while maintaining the same error, improves the sensitivity; alternatively, the sensitivity can remain constant while reducing the measurement time even further. Such capabilities could have significant impact on a broad range of measurement scenarios with large, time-varying signals, such as characterization of circuit performance, identifying transient biological signals, and more.

\section*{Acknowledgments}
This work was financially supported by the European Union’s Horizon 2020 research and innovation programme under FET-OPEN GA No.~828946–PATHOS. G.H.~also acknowledges support from the Melbourne research scholarship. S.M.~also acknowledges financial support from the PNRR MUR project PE0000023-NQSTI. N.B.~and F.C.~also acknowledge financial support by the European Commission’s Horizon Europe Framework Programme under the Research and Innovation Action GA No.~101070546–MUQUABIS. N.B.~also acknowledges financial support by the Carl Zeiss Stiftung (HYMMS wildcard), the Ministry of Science and Technology, Israel, the innovation authority (Project No.~70033), and the ISF (Grants No.~1380/21 and No.~3597/21). F.C.~also acknowledges financial support by the European Defence Agency under the project Q-LAMPS Contract No.~B~PRJ-RT-989.

\clearpage
\bibliography{mainV2}

\clearpage
\onecolumngrid
\appendix
\section*{Supplementary material}
\setcounter{table}{0}
\renewcommand{\thetable}{S\arabic{table}}
\setcounter{figure}{0}
\renewcommand{\thefigure}{S\arabic{figure}}

\begin{figure*}[b]
    \centering
    \subfloat[\label{fig:lfReal}]{%
        \centering
        \includegraphics{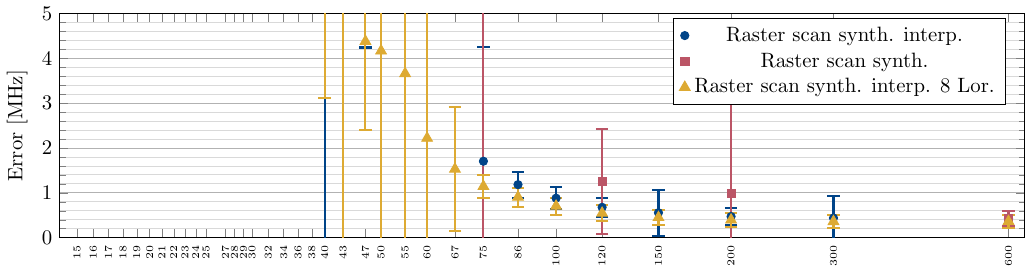}
    }

    \subfloat[\label{fig:nnSynth}]{%
        \centering
        \includegraphics{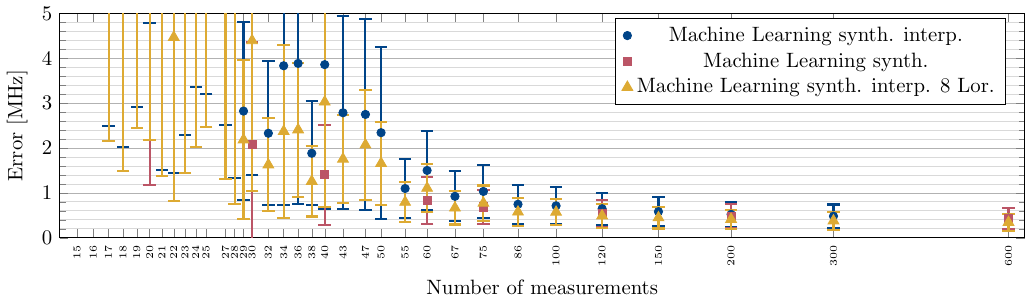}
    }
    \caption{\protect\subref{fig:lfReal} Normalised mean error of subsampled raster scans: blue circles show the error for $2\,000$ synthethic interpolated samples (the same of \cref{fig:nnlfIntSynth10k} in the main text). 
    In red squares, the same $2\,000$ synthetic samples that were subsampled but not interpolated.
    The yellow plot with triangles is similar to the blue, however all data samples are filtered to display eight non-overlapping Lorentzians.
    \protect\subref{fig:nnSynth} \ac{mae} of \ac{ml} networks:
    in red squares, networks were trained on $10\,000$ synthetic samples, and validated with $2\,000$ synthetic samples. 
    In the latter, the networks were trained for specific number of data points (trained on subsampled datasets), i.e, each point on the graph depicts the result of a different network. 
    The blue circles depict the result of one network, that was trained with $10\,000$ synthetic samples, and validated with $2\,000$ synthetic samples, all at full length with 600 measurement points (the same of \cref{fig:nnlfIntSynth10k} in the main text). The subsampled test data was linearly interpolated to the full length (600 points) before using the network to get the predictions.
    The yellow plot with triangles is similar to the blue, however it was trained on a dataset in which all data samples displayed eight non-overlapping Lorentzians.}
    \label{fig:nnlfRealInt}
\end{figure*}
In \cref{fig:lfReal} we report the normalized error for the raster scans of different simulated datasets and at different subsampling levels. The blue dots are calcolated on $2\,000$ samples that were subsampled to the specified number of measurements and then linearly interpolated to the full size of 600 points. The yellow triangles are generated similarly with the exception that the samples are filtered to always have eight resonances without overlapping. In that case, the error is understandably lower. Finally, the red squares are the normalized errors for $10\,000$ not interpolated samples.
\cref{fig:nnSynth} is equivalent to \cref{fig:lfReal} but reports the \ac{mae} of the trained \ac{ml} models. For the blue circles and yellow triangles the samples with less than 600 measurement points were linearly interpolated to the full size. Thus, in those cases we trained a single model with the full raster scan, and we tested it with the interpolated data. The red squares represent the results of the models trained directly on the subsampled data. Thus for every point of the curve, a different model was trained and tested. The models used for the yellow triangles in \cref{fig:nnSynth} are trained with the same data used for the yellow triangle of \cref{fig:lfReal}. Also in this case the error using the filtered data is understandably lower respect to the one that presents overlapping Lorentzians. However, it is remarkable that there is only a slightly difference between the two cases and that the \ac{ml} models, contrarily to the raster scan, performs well also with the unfiltered data.

\begin{figure}[th]
    \centering
    \includegraphics[width=0.5\linewidth]{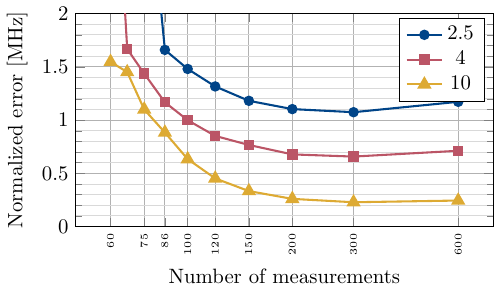}
    \caption{Normalized error of raster scanning on samples with \ac{snr} values of 2.5 (blue circles), 4 (red squares), and 10 (yellow triangles).}
    \label{fig:LFwidthtraining}
\end{figure}
In \cref{fig:LFwidthtraining} we report the normalized error of raster scanning on the same samples used for the tests of \cref{fig:10mhz,fig:widthrange} of the main paper. They are also the same values used in \cref{fig:rastersnr} where we report the not normalized error (the average of the values only where the raster scanning is able to find all the 8 resonances).

\begin{figure*}[h]
    \centering
    \subfloat[\label{fig:realTest1}]{%
        \centering
        \includegraphics{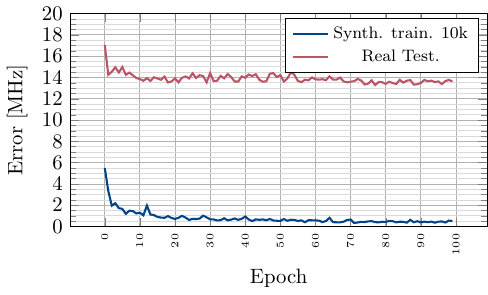}
    }
    \subfloat[\label{fig:realTest2}]{%
        \centering
        \includegraphics{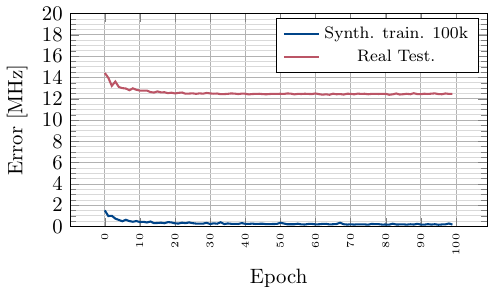}
    }
    \caption{Learning curves for the models shown in \cref{fig:trainSynth} of the main text, trained with $10\,000$ \protect\subref{fig:realTest1} and $100\,000$ \protect\subref{fig:realTest2} synthetic samples. For both models, the plots show the evolution of the error on the synthetic training set and on the same real test set used in \cref{fig:10k} of the manuscript.}
    \label{fig:realTest}
\end{figure*}

Finally, in \cref{fig:realTest}, we present the evolution of the error on both the training set and a real test set for the models shown in \cref{fig:trainSynth} of the main text. \cref{fig:realTest1} shows the training progress of the model trained with $10\,000$ synthetic samples, while \cref{fig:realTest2} corresponds to the model trained with $100\,000$ samples. In both cases, the training error (along with the validation error shown in \cref{fig:trainSynth}) decreases to below $0.5$ MHz. However, the test error on real data remains above $12$ MHz, indicating significant overfitting to the synthetic data.

\end{document}